\documentclass{article}

\usepackage[margin=1.4in,headsep=10pt]{geometry}
\usepackage{amsmath}
\usepackage{amssymb}
\usepackage{amsfonts} 
\usepackage{color}
\usepackage{graphicx}

\title{Generating theorem proving procedures from axioms of Truncated Predicate Calculus}
\author{Grzegorz Wiaderek \\
	Infoster Sp. z o.o. \\ \texttt{gwiaderek@infoster.com.pl} \\ Iwona Skalna \\ AGH Unversity of Science and Technology}
\date{}
\begin{document}
	\maketitle
	\begin{abstract}
		We present a~novel approach to the problem of automated theorem proving. Polynomial cost procedures that recognise sentences belonging to a theory are generated on a basis of a set of axioms of the so-called Truncated Predicate Calculus being a~subset of standard predicate calculus. Several exemplary problems are included to show the performance of the proposed approach.
	\end{abstract}


	\section{Introduction}
	In this paper we present a~novel approach to the problem of automated theorem proving. It is well known that the main problem of automated theorem proving is the time complexity. A~basic algorithm for predicate calculus is based on the resolution rule \cite{ChangLee:SLMTP:1973}. Theoretically, it can prove any theorem that can be proved, but its cost grows exponentially with the length of the shortest proof. This causes that it is almost impossible to obtain the proof for more complex problems. Unfortunately, in the general case a~better algorithm simply does not exist, and searching it would be an attempt to invent a computer perpetuum mobile. On the other hand, for a~specific problem encoded by a~set of axioms, there might exist a~procedure with polynomial complexity. Thus, instead of laborious search of the entire space of proofs, one could first automatically find such procedure, and then use it to prove the theorem. This is the main idea of the approach proposed in this paper.
	
	Obviously, this approach cannot be applied to an arbitrary theory (understood as a~set of axioms that encodes a (computer) problem). It is important to be aware that we will not deal with problems of the rank of ``real'' mathematical theories, because, as is well known, they are generally undecidable. Therefore, the potential area of interest will be, at least at the beginning, to create an improved version of the programming language Prolog. The lofty assumption of the developers of Prolog was to allow programmers to focus on what is to be calculated and not on how it has to be calculated. Alas, at all the undoubted usefulness of this language, this assumption can hardly be considered to be realized, since the correct formulation of axioms does not guarantee the correctness of the program, not mentioning about its effectiveness. Thus, if it would be possible, for example, to settle for defining a~theory of sorting in the form of axioms, and a~program would be able to generate from those axioms a~sorting procedure, this would be a~significant progress. An attempt to obtain such a~progress is described in this paper.

This article is based on the results of an experimental program, which (like most of existing programs) is still in the development phase. We are far from saying that we have already fully explored the subject, but what has been achieved seems to be interesting and promising. We will try to present here some general high-level approach to the problem. It is elegant and works, although in practice only partially due to the low-level technical problems.

A few words about notation. For the sake of maximum readability, we wrote the sample procedures in the pseudo-code mixing the lambda-calculus \cite{Barendregt:LCT:1993} with the traditional mathematical notation. In the real system, all procedures are written in a lisp-like language based on a~pure lambda-calculus, which is unfortunately not very readable.

\section{Basic concepts}

\subsection{Truncated Predicate Calculus}
\label{sec:tcp}
We begin from defining our basic system, i.e., the system that we will use to write theorems, axioms and proofs. Because the problem of theorem proving is very difficult, we will make, at least at the beginning, some necessary simplifications, they will be described in what follows. The basis of our system is the standard predicate calculus \cite{ChangLee:SLMTP:1973}. It may seem far too simple to perform real proofs, but for the purposes of our approach it is even too complex. It will be, therefore, truncated as much as possible, but so as to preserve its ``basic nature''. The resulting system will be called Truncated Predicate Calculus (TPC). It must be, however, emphasized that this will be something lying somewhere between the standard predicate calculus and a formal grammar. Intuitively speaking, any superfluous ballast will be removed.

The purpose of the first simplification is to obtain a linear structure of a proof, more resembling a~derivation from a formal grammar than a traditional proof in the predicate calculus (having in fact a graph structure). We want our theory to consist of a set of axioms resembling productions, i.e., in the form of clauses p$\rightarrow$q, and a starting sentence s. A~proof will now have a form of a sequence of sentences: the sentence s will be the first in this sequence, whereas each next sentence will result from the previous one by applying one of the axioms. The last sentence will be the proved theorem. In order to obtain this, we must not distinguish between functions and predicates, get rid of the explicit support for logical operators, slightly modify input theories and add axioms of logic. The following exemplary Prolog-style problem illustrates this idea:\\

\noindent p1: Parent(Adam, John) \\
p2: Parent(John, Peter) \\
p3: Parent(Peter, Olga)\\
a1: Parent(p, c) $\rightarrow$ Ancestor(p, c) \\
a2: (Parent(p, a) $\wedge$ Ancestor(a, o)) $\rightarrow$ Ancestor(p, o)\\
goal: Ancestor(Adam, Olga)\\

\noindent In order to fit the above problem into TPC, we must transform it into the following form: \\

\noindent s: S \\
p1: x $\rightarrow$ And(Parent(Adam, John), x) \\
p2: x $\rightarrow$ And(Parent(John, Peter),  x) \\
p3: x $\rightarrow$ And(Parent(Peter, Olga),  x) \\
a1: And(Parent(p, c), x) $\rightarrow$ And(Ancestor(p, c), x) \\
a2: And(Parent(p, a), And(Ancestor(a, o) , x)) $\rightarrow$ And(Ancestor(p, o), x) \\

\noindent and add axioms of logic: \\

\noindent l1: And(x, y) $\rightarrow$ x \\
l2: And(x, y) $\rightarrow$ y \\

\noindent Now we can prove the goal clause, the proof is: s, p3, a1, p2, a2, p1, a2, l1.

\paragraph{Remark.} Generally, the above approach is legitimate under the assumption that, similarly as in a formal grammar, axioms and the starting sentence are relatively simple, and only the goal sentence can be arbitrary long.\\

The second simplification concerns restrictions on the occurrence of variables. Basically, variables should not appear in sentences. This means that the starting sentence should not contain any variable, and axioms should be formulated so that they do not introduce any new variable to the resulting sentence. One might ask why this simplification is needed. Generally, the subject of interest are the sets of sentences. Sentences constructed only from constants have simple and uniform structure. Moreover, there will be no confusions, which may happen when single objects and sets of objects are put in the same ``bag''. For example, the sentence P(F(x)) is, in some sense, the set of all concrete sentences like P(F(A)).

\subsection{Iterative schemes and multi-indexes}
\label{sec:iter_scheme}
Our main goal is to obtain a function of type T$\rightarrow$B with polynomial cost (with respect to the size of the goal tree), which will return True for all the trees of the theory (and only for them). Assume that the theory consists of the start tree S and $n$ axioms A$_i$. It might seem natural to try constructing this function iteratively in $n$ steps so that in every step we would get a~function of type T$\rightarrow$B for S and first $k$ axioms. Unfortunately, this approach will probably fail. This is because functions of type T$\rightarrow$B seem to be too rigid to be used in the next steps of iteration process. 
Therefore, we will use functions of type T$\rightarrow$T$\rightarrow$B instead. Let us define a~characteristic function of the set of axioms as a~function of type T$\rightarrow$T$\rightarrow$B that will return a~set of all trees d that can be reached with a sequence of axioms starting from a~given start tree t. We will construct such functions iteratively, adding one axiom in each step, and finally we will apply the resulting function to the S tree to obtain a function of type T$\rightarrow$B.

Our axioms are clauses in the form of p$\rightarrow$q, so in a~natural way we can identify them with functions of type T$\rightarrow$T, which for a given tree t return any result tree d if the axiom matches the tree t, or the empty tree Null otherwise. However, since we want to operate on the relations of type T$\rightarrow$T$\rightarrow$B, an axiom a: T$\rightarrow$T will be from now on identified with the relation: $\lambda$t:T.$\lambda$d:T.(a(t)=d).

We will now define several operators on functions of type T$\rightarrow$T$\rightarrow$B. Let $\alpha$, $\beta$ be a~pair of functions of type T$\rightarrow$T$\rightarrow$B, then:
\[
\alpha\cdot\beta:=\lambda\mathrm{t:T}.\bigcup_{\mathrm{d}\in\alpha(\mathrm{t})}\beta(\mathrm{d}).
\]
By $\alpha^n$, we mean the recursively defined operation:
\[
\alpha^n:=
\left\{\
\begin{array}{ll}
\varepsilon, & n=0, \\
\alpha^{n-1}\cdot\alpha, & n>0,
\end{array}
\right.
\]
where $\varepsilon:=\lambda\mathrm{t}:\mathrm{T}.\lambda\mathrm{d}:\mathrm{T}.(\mathrm{t}=\mathrm{d})$,
\[
\alpha^*:=\lambda\mathrm{t}:\mathrm{T}.\bigcup_{i=0}^{\infty}\alpha^i(\mathrm{t}),
\]
\[
\alpha\mid\beta:=\lambda\mathrm{t}:\mathrm{T}.\alpha(\mathrm{t})\cup\beta(\mathrm{t})
\]
Let us add, for the sake of clarity, that the dot operator is associative. Additionally, let us assume that the operator * has the highest priority and the operator $\mid$ the lowest.

The above definitions formally define the semantics of expressions composed of axioms and operators ${*,\cdot,\mid}$. They will be called iterative expressions. An iterative expression that contains only $\cdot$ operators will be called a~\textit{specific expression}. A specific expression can be easily reduced to a~single TPC clause, which we identify with the T$\rightarrow$T$\rightarrow$B relation in a similar way as in the case of axioms. Now we will look at iterative expressions in a different way, i.e., as if they were iterative schemes. We will start from defining a set of multi-indexes. The set of multi-indexes M contains natural numbers (N) and lists of any multi-indexes (we will write them in curly brackets separated by commas). For example: 7, \{1, 2, 3\}, and \{8, 0, \{4, 6, 1\}, 2\}. The length of a multi-index $m$, denoted by $|m|$, is defined as follows: if $m$ is a natural number, then $|m|$ equals to that number, otherwise, $|m|$ equals to the length of the list. So, in the above example the lengths of multi-indexes are: 7, 3, and 4. We will also introduce the operator of picking the nth element of the list, which will be denoted with square brackets. For example: \{1, 2, 3\}[2] = 2, \{8, 0, \{4, 6, 1\}, 2\}[3] = \{4, 6, 1\}.

By the \textit{iterative scheme} of an iterative expression, we mean a~function of M$\rightarrow$T$\rightarrow$T$\rightarrow$B type, which, for a~given multi-index $m$, returns a specific iterative expression, which can be defined recursively as follows:
\begin{itemize}
	\item The value of the expression $a(m)$, where $a$ is an axiom is $a$,
	\item The value of the expression $\alpha^*(m)$ is $\alpha(\mathrm{m}[1])\cdot\alpha(\mathrm{m}[2])\cdot\ldots\cdot\alpha(\mathrm{m}[|\mathrm{m}|])$,
	\item The value of the expression  $(\alpha_1\cdot\alpha_2\cdot\ldots\cdot\alpha_j)(m)$, in the case when $\alpha_1$ is an axiom, is $\alpha_1\cdot((\alpha_2\cdot\ldots\cdot\alpha_j)(m))$, otherwise it is $\alpha_1(m[1])\cdot((\alpha_2\cdot\ldots\cdot\alpha_j)(\textrm{Tail}(m)))$,
	\item The value of the expression $(\alpha_1|\alpha_2|\ldots|\alpha_j)(m)$ is $\alpha_{m[1]}(m[2])$, where $m$ must have the length of 2.
\end{itemize}
This above (somewhat complex) definition enables us to obtain a~simple and intuitive way of iterating through all specific expressions of a given iterative expression (that is, our space of proofs). For example:\\

\noindent (a$^*\cdot$b)$^*\cdot$a$^*(\{\{2,0,1\},3\})= \  $a$\cdot$a$\cdot$b$\cdot$b$\cdot$a$\cdot$b$\cdot$a$\cdot$a$\cdot$a.\\

We will now specify how to iteratively construct the procedure T$\rightarrow$T$\rightarrow$B which is the equivalent of a given set of axioms. For example, for a set of three axioms {a, b, c}, in the first step we create the procedure a$^*$, in the second step (a$^*\cdot$b)$^*\cdot $a$^*$ and in the third step ((a$^*\cdot$b)$^*\cdot$a$^*\cdot$c)$^*\cdot$ (a$^*\cdot$b)$^*\cdot $a$^*$. In general: if $\alpha$ is the procedure obtained in $(n-1)$ step, and $\mathrm{a}_n$ is the $n$th axiom, then in the $n$th step we create the procedure $(\alpha\cdot\mathrm{a}_n)^*\cdot\alpha$. It is easy to show that in this way we obtain the entire space of proofs.
 
From now on, for the simplicity of notation, we will identify a multi-index with its length whenever the argument of type N is required.

\subsection{Splitting to atoms}
\label{sec:split_atoms}
Trees have a complex structure, which makes it difficult to describe a~relation on the set of trees directly. Therefore, we split such relation into \textit{atoms}\footnote{These atoms should not be confused with Prolog atoms.}. A single atom consists of a pair of paths in a tree. Each path has a~linear structure, slightly resembling the structure of a~regular expression, so handling it is relatively simple.

More precisely: we define a path as a T$\rightarrow$T function transforming a tree into one of its subtrees. The basic building blocks for paths are TPC clauses in the form of p$\rightarrow$v, where v is a single variable; we will write them in square brackets. For example, the axiom\\

\noindent a: P(R(x, z), y)$\rightarrow$P(x, R(y, z))\\

\noindent is split into the intersection of three atoms:\\

\noindent Intersect(\\%
\hspace*{20pt}EqualsLR(\\%
\hspace*{40pt}[P(R(x, y), z)$\rightarrow$x],\\%
\hspace*{40pt}[P(x, y)$\rightarrow$x]),\\%

\noindent\hspace*{20pt}EqualsLR(\\%
\hspace*{40pt}[P(x, y)$\rightarrow$y],\\%
\hspace*{40pt}[P(x, R(y, z))$\rightarrow$y]),\\%

\noindent\hspace*{20pt}EqualsLR(\\%
\hspace*{40pt}[P(R(x, z), y)$\rightarrow$z],\\%
\hspace*{40pt}[P(x, R(y, z))$\rightarrow$z]) \\%
) \\

\noindent The semantics of the EqualsLR operator of type ((T$\rightarrow$T)$\times$(T$\rightarrow$T))$\rightarrow$(T$\rightarrow$T$\rightarrow$B) is as follows: it takes two paths and returns a T$\rightarrow$T$\rightarrow$B function that applies these paths to its own arguments and returns True if both return the same. So far, we have only used atoms based on the EqualsLR operator. Ultimately (in the future), we will need two additional operators: EqualsLL and EqualsRR, which will test the equality of subtrees in the left and right tree, respectively.	
The basic operation performed on paths is the combination of functions marked with the dot. For example: [P(x,y)$\rightarrow$x]$\cdot$[R(x,y)$\rightarrow$y] = [P(R(x, y), z)$\rightarrow$y]. The n-fold repetition, in turn, indicated by the upper index, is defined as follows: $\omega^0:=[$x$\rightarrow$x$]$, $\omega^{n+1}:=\omega^n\cdot\omega$. For example, [F(x)$\rightarrow$x$]^4=[$F(F(F(F(x))))$\rightarrow$x$]$.
At further stages of processing, we will obtain atoms with a~more complex structure that is a function of natural and multi-index parameters.

\section{Architecture of system}
\label{sec:architecture}
The architecture of our system is based on five main solvers:
\begin{itemize}
	\item Sigma-Solver,
	\item Inclu-Solver,
	\item Math-Solver,
	\item Delta-Solver,
	\item Final-Solver.
\end{itemize}
They will be described in the following sections.

\subsection{Sigma-Solver and idea of function described by formula}
\label{sec:sigma_solver}
Sigma-Solver generates for the given iterative scheme the function M$\rightarrow$T$\rightarrow$T$\rightarrow$B, which for a given multi-index returns the procedure T$\rightarrow$T$\rightarrow$B corresponding to a given combination of axioms. The task formulated in this way, can be realized in an easy manner, but the function generated by Sigma-Solver gives us something more. Imagine an analytic function described by the formula and the same function in the form of a black box. If we only want to get values of this function for given arguments, both forms are equally useful. However, having the formula, we can go further: we can analyze this function to find its extrema, zeros, etc. A~function generated by Sigma-Solver is such a~formula for our relation on the set of trees. It can be further analyzed, as we will show later. Let's look at the exemplary set of the following two axioms:\\

\noindent a: P(R(x, z), y) $\rightarrow$ P(R(x, F(z)), y) \\
b: P(R(x, z), y) $\rightarrow$ P(x, R(y, z)) \\

What can we get from the iterative scheme (a*$\cdot$b)*? Assume that we start from a~tree t, for example the left-hand one presented in Fig. \ref{fig:fig1} (for better readability we present the trees graphically):
\begin{figure}[!htb]
	\centering
	\includegraphics[width=0.35\textwidth]{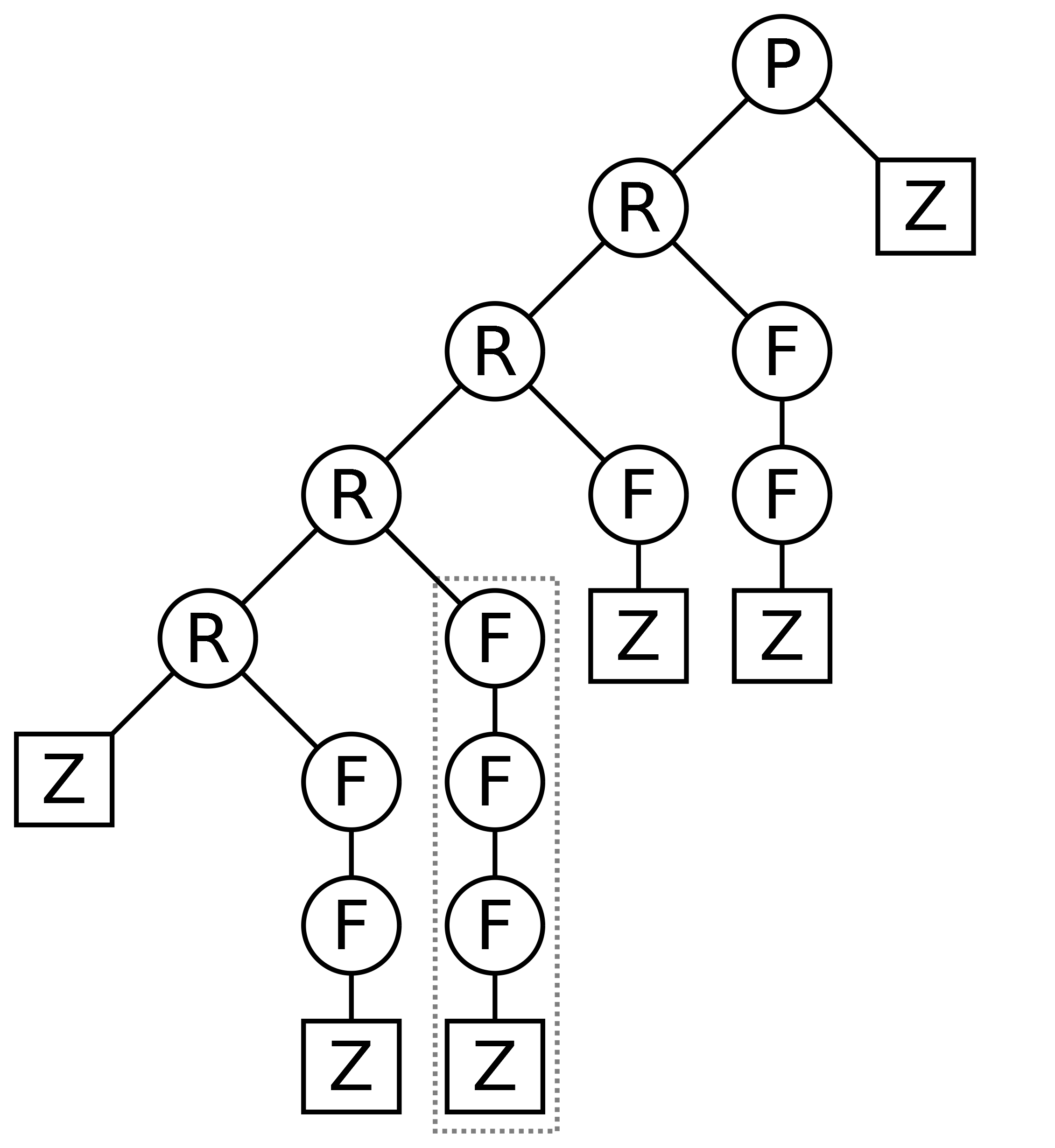}\quad\quad\quad
	\includegraphics[width=0.35\textwidth]{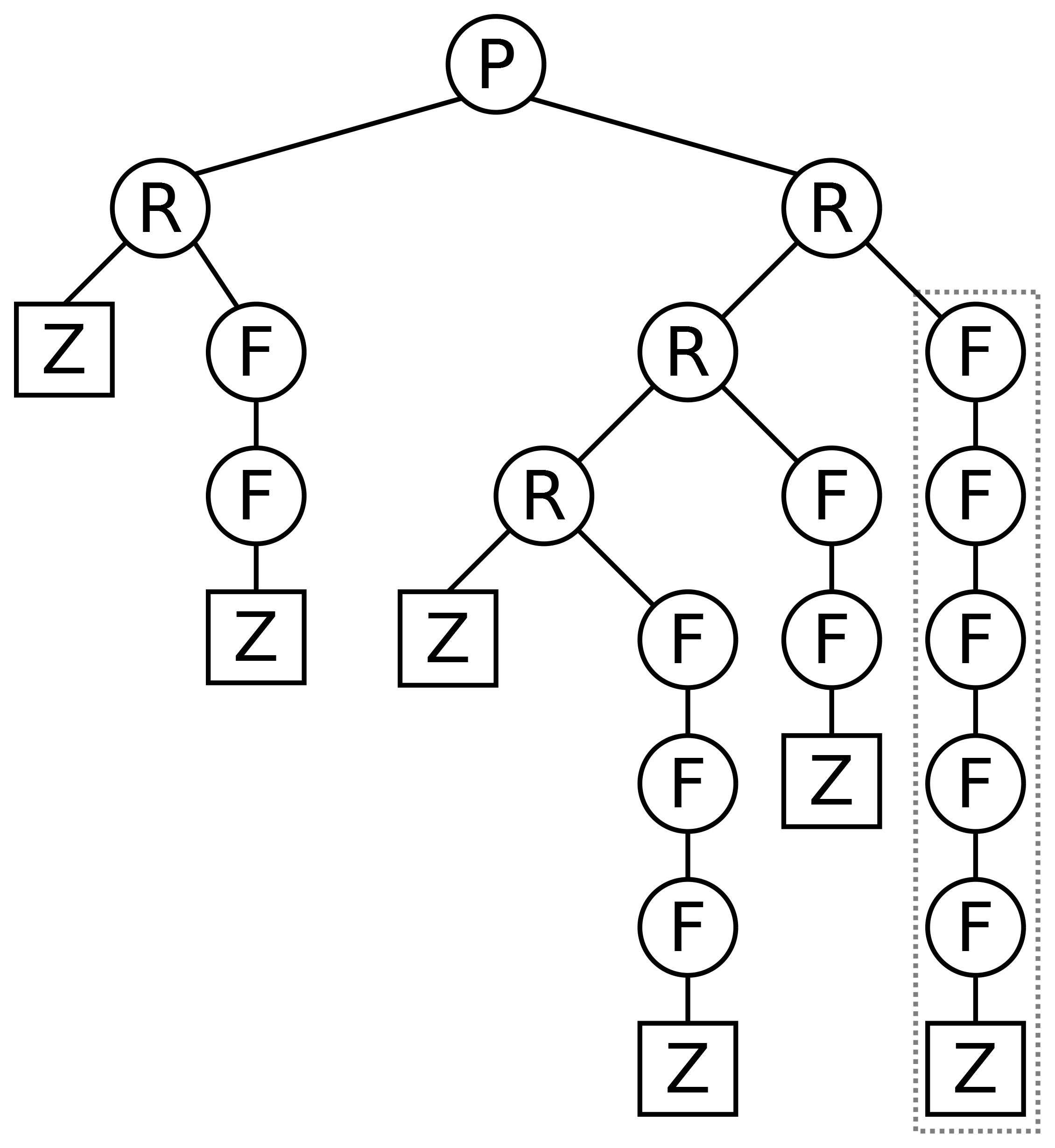}
	\caption{Examples of start and goal trees}
	\label{fig:fig1}
\end{figure}

Let $k$ denote the number of R nodes in the leftmost branch (for the considered tree $k = 4$). It can be easily seen that b axiom can be used 0,$\ldots$,$k$ times. Thus, we can divide the set of goal trees into $k+1$ disjoint subsets Z$_i$ ($i=0,\ldots,k$). The goal tree d presented in Fig.~\ref{fig:fig1} on the right belongs to Z$_3$. The trees in the subset Z$_i$ must satisfy the following condition: the number of symbols F in the ``corresponding'' (as those marked in Fig. \ref{fig:fig1} with the gray borders) branches of t and d must be in the ``less or equal'' relation. 

The function generated by Sigma-Solver for the (a*$\cdot$b)* scheme is shown below:\\

\noindent $\lambda m$:M.Intersect(\\[6pt]
\hspace*{20pt}EqualsLR(\\
\hspace*{40pt}[P(x, y)$\rightarrow$x]$\cdot$[R(x, y)$\rightarrow$x]$^m$,\\
\hspace*{40pt}[P(x, y)$\rightarrow$x]),\\[6pt]
\noindent\hspace*{20pt}EqualsLR(\\
\hspace*{40pt}[P(x, y)$\rightarrow$y],\\
\hspace*{40pt}[P(x, y)$\rightarrow$y]$\cdot$[R(x, y)$\rightarrow$x]$^m$),\\[6pt]
\noindent\hspace*{20pt}IterIntersect(\\
\hspace*{40pt}$\lambda i$:N.EqualsLR(\\
\hspace*{60pt}[P(x, y)$\rightarrow$x]$\cdot$[R(x, y)$\rightarrow$x]$^{i-1}\cdot$[R(x, y)$\rightarrow$y], \\
\hspace*{60pt}[P(x, y)$\rightarrow$y]$\cdot$[R(x, y)$\rightarrow$x]$^{m-i}\cdot$[R(x, y)$\rightarrow$y]$\cdot$[F(x)$\rightarrow$x]$^{m[i]}$),
1, $m$)\\
) \\

\noindent where
\[
\textrm{IterIntersect}(\mathrm{f},\mathrm{u},\mathrm{v})=\bigcap_{i=\mathrm{u}}^\mathrm{v}\mathrm{f}(i).
\]

\subsection{Inclu-Solver and Math-Solver}
\label{sec:inclsol_mathsol}
Let's take two functions f, g of type M$\rightarrow$T$\rightarrow$T$\rightarrow$B. The task of Inclu-Solver is to generate a~function h of type M$\rightarrow$M$\rightarrow$B, which will return True if a~pair of multi-indexes ($m, n$) satisfies f($m$)$\subset$g($n$). Let's look at an exemplary pair of axioms:\\

\noindent a: P(x, y)$\rightarrow$P(F(F(x)), G(y))\\
b: P(x, y)$\rightarrow$P(F(x), G(y))\\

Now suppose that we are interested in the following inclusion of the iterative schemes: a$\cdot$b$\cdot$a$^*\subset$ b$\cdot$b$\cdot$a$^*\cdot$b. First, Sigma-Solver generates the function f for the a$\cdot$b$\cdot$a$^*\cdot$b scheme:\\

\noindent $\lambda n$:M.Intersect(\\[6pt]
\hspace*{20pt}EqualsLR(\\
\hspace*{40pt}[P(x, y)$\rightarrow$x],\\
\hspace*{40pt}[P(F(F(F(x))), y)$\rightarrow$x]$\cdot$[F(F(x))$\rightarrow$x]$^n\cdot$[F(x)$\rightarrow$x]), \\
\hspace*{20pt}EqualsLR(\\
\hspace*{40pt}[P(x, y)$\rightarrow$y], \\
\hspace*{40pt}[P(x, G(G(y)))$\rightarrow$y]$\cdot$[G(x)$\rightarrow$x]$^{n+1}$)\\
)\\

\noindent and the function g for the b$\cdot$a$^*\cdot$b scheme:\\

\noindent $\lambda k$:M.Intersect(\\[6pt]
\hspace*{20pt}EqualsLR(\\
\hspace*{40pt}[P(x, y)$\rightarrow$x],\\
\hspace*{40pt}[P(F(x), y)$\rightarrow$x]$\cdot$[F(F(x))$\rightarrow$x]$^k\cdot$[F(x)$\rightarrow$x]),\\
\hspace*{20pt}EqualsLR(\\
\hspace*{40pt}[P(x, y)$\rightarrow$y],\\
\hspace*{40pt}[P(x, G(y))$\rightarrow$y]$\cdot$[G(x)$\rightarrow$x]$^{k+1}$)\\
)\\

\noindent Then, Inclu-Solver in response to the query:\\

\noindent INCLUDES(f, g) \\

\noindent will generate the function h: \\

\noindent $\lambda n$:M.$\lambda k$:M.And( \\
\hspace*{10pt}$2n+3=2+1$,\\
\hspace*{10pt}$n+2=k+1$\\
)\\

Thanks to the splitting into atoms having the linear structure, this task can be done relatively easily. It should be underlined that at this stage we get rid of all operations on trees. What remains is a~set of equations (and, in a general case, also inequalities) that can be dealt with traditional mathematical methods. In our system such methods are included in Math-Solver.

We now want to find the set of natural numbers $n$ for which there are natural numbers $k$ that satisfy the above system of equations. For this purpose, Math-Solver must solve the above (over-determined) system of equations with respect to $k$. The solution is: $k=n+1$. We must add a condition $k\geqslant0$, which gives an additional condition $n\geqslant -1$. Since this condition is satisfied for each $n\in$ N, therefore we have just proved the complete inclusion a$\cdot$b$\cdot$a$^*\cdot$b $\subset$ b$\cdot$a$^*\cdot$b. Note that the additional condition can be important, e.g., if we invert our original query, i.e., we will consider the query b$\cdot$a$^*\cdot$b $\subset$ a$\cdot$b$\cdot$a$^*\cdot$b, we would get as a~result the interval $[1,\infty)$ instead of the whole set of natural numbers, as b$\cdot$b $\not\subset$ a$\cdot$b$\cdot$a$^*\cdot$b.

\subsection{Repetitive component of unraveling system of axioms and Delta-Solver}
\label{sec:deltasol}
Basically, the unraveling of every system of axioms involves two independent components, let's call them a repetitive component and an unambiguous component. In this section, we will focus on the first one and in the next section on the second one.

The repetitive component is related to the repetitions occurring in the iterative scheme (i.e., the same trees obtained for different multi-indexes). By detecting such repetitions, it is often possible to reduce the iterative scheme to a simpler one. And this is the Delta-Solver task. In fact, Delta-Solver is a high-level object that uses the three lower-level solvers mentioned earlier: Sigma-Solver, Inclu-Solver and Math-Solver.

\begin{figure}[!htb]
	\centering
	\includegraphics[width=0.15\textwidth]{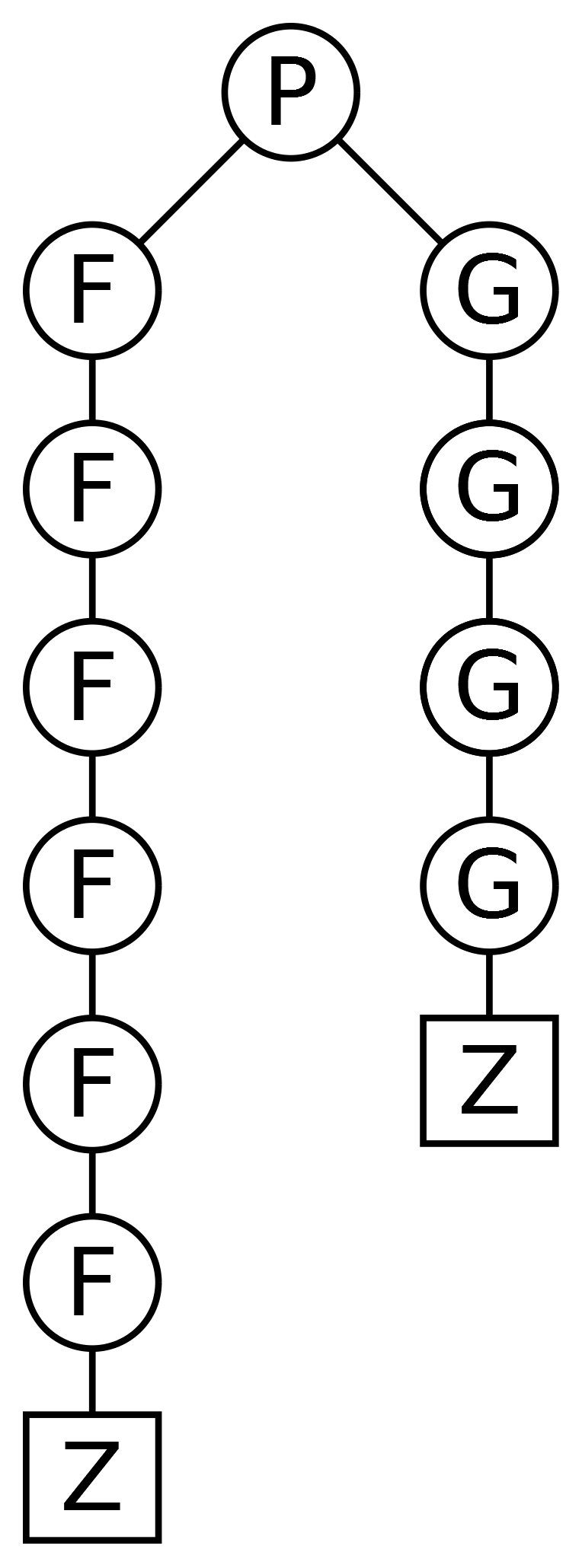}
	\caption{Example of goal tree}
	\label{fig:fig2}
\end{figure}

Let us return to the exemplary set of axioms from Section \ref{sec:inclsol_mathsol}. It is easy to see that starting from the tree P(Z,Z), we can generate trees such as the one presented in Fig.~\ref{fig:fig2}, wherein the number of F's in the left subtree must be in the range $[n,2n]$, where $n$ is the number of G's in the right subtree. The iteration scheme (a$^*\cdot$b)$^*\cdot$a$^*$ contains repetitions and can be reduced to b$^*\cdot$a$^*$. How to achieve this? The crucial thing is to reduce the (a$^*\cdot$b)$^*$ scheme. Delta-Solver analyses subsequent elements of the sequence (a$^*\cdot$b)$^*$, i.e., (a$^*\cdot$b)$^1$, (a$^*\cdot$b)$^2$, etc. The (a$^*\cdot$b)$^1$ scheme cannot be simplified. The (a$^*\cdot$b)$^2$ equals a$^*\cdot$b$\cdot$a$^*\cdot$b. Analyzing the following elements: a$^0\cdot$b$\cdot$a$^*\cdot$b and a$^1\cdot$b$\cdot$a$^*\cdot$b it can be observed that a$^1\cdot$b$\cdot$a$^*\cdot$b $\subset$ a$^0\cdot$b$\cdot$a$^*\cdot$b (which is exactly what we have done in Section \ref{sec:inclsol_mathsol}). And this allows us to reduce the whole (a$^*\cdot$b)$^*$ scheme to the form b$^*\cdot$(a$^*\cdot$b)$|\varepsilon$. So the scheme (a$^*\cdot$b)$^*\cdot$a$^*$ reduces to b$^*\cdot$a$^*\cdot$b$\cdot$a$^*|$a$^*$. The b$^*\cdot$a$^*\cdot$b$\cdot$a* scheme can be reduced to b$^*\cdot$b$\cdot$a$^*$, therefore the whole scheme b$^*\cdot$a$^*\cdot$b$\cdot$a$^*|$a$^*$ reduces to b$^*\cdot$b$\cdot$a$^*|$a$^*$, which is equal to b$^*\cdot$a$^*$.

\subsection{Unambiguous component of unraveling system of axioms and Final-Solver}
The unambiguous component of unraveling the system of axioms is related to finding an ``unambiguous'' atom, i.e., an atom that can be matched only in a single way for a~given pair of trees. Let's start with a simple example: consider the following single axiom:\\

\noindent a: P(x)$\rightarrow$P(F(x))\\

\noindent For the a* scheme Sigma-Solver generates the function: \\

\noindent $\lambda j$:M.EqualsLR(\\
\hspace*{20pt}[P(x)$\rightarrow$x],\\
\hspace*{20pt}[P(x)$\rightarrow$x]$\cdot$[F(x)$\rightarrow$x]$^j$ \\
) \\

\noindent For a particular pair of trees, for example t = P(F(Z)), d = P(F(F(F(F(Z))))))), we can ``tune'' the above function with a~linear cost, by simply testing it for $j = 0,1,2\ldots$, until we find a~match (as here for $j$ = 3) or until the path [P(x)$\rightarrow$x]$\cdot$[F(x)$\rightarrow$x]$^j$ stops to match the tree d.

Let's now go back to the example from Section \ref{sec:inclsol_mathsol}, where Sigma-Solver for the b$^*\cdot$a$^*$ scheme generated the function:\\

\noindent $\lambda n$:M.$\lambda k$:M.Intersect(\\[6pt]
\hspace*{20pt}EqualsLR(\\
\hspace*{40pt}[P(x, y)$\rightarrow$x],\\
\hspace*{40pt}[P(x, y)$\rightarrow$x]$\cdot$[F(x)$\rightarrow$x]$^{2n+k}$),\\[6pt]
\hspace*{20pt}EqualsLR(\\
\hspace*{40pt}[P(x, y)$\rightarrow$y], \\
\hspace*{40pt}[P(x, y)$\rightarrow$y]$\cdot$[G(y)$\rightarrow$y]$^{n+k}$)\\
)

\noindent which is the intersection of two atoms. For a specific pair of trees, for example t=P(Z,Z), d=P(F(F(F(F(F(F(F(Z))))))))), G(G(G(G(G(Z)))))), tuning of the first atom results in the equation: $n+2k=8$, and tuning of the second atom gives the equation: $n+k=5$. This system of 2 equations with 2 unknowns is complemented with two conditions $n\geqslant0$, $k\geqslant0$, and the final solution is: $n=2$, $k=3$.

Now let's look once again at the function from Section \ref{sec:sigma_solver}. For the exemplary t and d trees from Section \ref{sec:sigma_solver}, tuning of both the first and the second atom will result in $m=3$. After substituting this solution into the third, iterated atom, we can also tune it, obtaining: $m[1]=1$, $m[2]=1$, and $m[3]=2$.

The above-described actions are performed in our model by a separate solver called Final-Solver (which, however, employs Math-Solver).

\subsection{Final algorithm}
\label{sec:final_algo}
The current version of the main algorithm is as follows: 
\begin{enumerate}
	\item We begin by simplifying (as much as possible) the original iterative scheme using Delta-Solver. We do it iteratively, adding 1 axiom in each step, in accordance with the idea described in Section \ref{sec:iter_scheme}.
	\item For the obtained iterative scheme, we generate the M$\rightarrow$T$\rightarrow$T$\rightarrow$B function using Sigma-Solver.
	\item Finally, we use Final-Solver to get the final T$\rightarrow$T$\rightarrow$B procedure.
\end{enumerate}

The main advantage of this algorithm is its simplicity. Nevertheless, it also has some drawbacks that will have to be dealt with in the future. We will return to this issue in Section~\ref{sec:future_work}. 

\section{Additional issues}

\subsection{Modular equations}
\label{sec:modulareq}
In order to make the view more complete, it is worth mentioning the often obtained modular equations. Here is a simple example:\\

\noindent a: P(x)$\rightarrow$P(F(x)) \\
b: P(x)$\rightarrow$P(F(F(x))) \\

\noindent Let's assume that we want to test the inclusion: a$^*\subset$ b$^*$. Sigma-Solver, for the a$^*$ and b$^*$ schemes, generates two procedures of type M$\rightarrow$T$\rightarrow$T$\rightarrow$B: \\

\noindent a$^*=\lambda n$:M.EqualsLR(\\
\hspace*{20pt}[P(x)$\rightarrow$x], \\
\hspace*{20pt}[P(x)$\rightarrow$x]$\cdot$[F(x)$\rightarrow$x]$^n$\\
) \\[6pt]

\noindent b$^*=\lambda k$:M.EqualsLR(\\
\hspace*{20pt}[P(x)$\rightarrow$x],\\
\hspace*{20pt}[P(x)$\rightarrow$x]$\cdot$[F(F(x))$\rightarrow$x]$^k$\\
) \\

\noindent Then, Inclu-Solver in response to the query: INCLUDES(a$^*$,b$^*$) generates the function:\\

\noindent f:M$\rightarrow$M$\rightarrow$B = $\lambda n$:M.$\lambda k$:M.($n=2k$)\\

\noindent Math-Solver solves the above equation and gives the set of natural numbers $n$ for which exists a~natural number $k$ that satisfies this equation, namely: \\

\noindent g:M$\rightarrow$B = $\lambda n$:M.($n\mod2=0$)

\subsection{More complex example}
\label{sec:more_complex}
Let us now consider the example from Section \ref{sec:sigma_solver} with an additional axiom c: \\

\noindent a: P(R(x, z), y)$\rightarrow$P(R(x, F(z)), y) \\
b: P(R(x, z), y)$\rightarrow$P(x, R(y, z)) \\
c: P(x, R(y, z))$\rightarrow$P(R(x, z), y) \\

How to solve this problem? The procedure, let's call it $\alpha$, generated by Sigma-Solver for the (a$^*\cdot$b)$^*\cdot$a$^*$ scheme (quite similar to the procedure for the (a$^*\cdot$b)$^*$ scheme presented previously) has no repetitions that could be reduced. The next procedure is ($\alpha\cdot$c)$^*\cdot\alpha$. Delta-Solver tries to reduce the ($\alpha\cdot$c)$^*$ scheme. In order to achieve this, it first tries to reduce the ($\alpha\cdot$c)$\cdot$($\alpha\cdot$c) scheme. 
From running the query: INCLUDES(($\alpha\cdot$c)$\cdot$($\alpha\cdot$c), ($\alpha\cdot$c)), the function f:~M$\rightarrow$M$\rightarrow$B is obtained. This function returns True for a pair of multi-indexes $(m, u)$, if a conjunction of the following conditions holds true\footnote{Unlike in the previous examples, this is only the function semantics. The actual function syntax generated by the program is much more complex. At the moment, the program doesn't simplify the function so nicely, although it is possible.}:\\

\noindent $m[1] \geqslant 1$ \\
$m[2] \geqslant 1$ \\
$m[1] + m[2] - u - 2 = 0$ \\
$m[1][i] - u[i] = 0$, for $i=1\ldots(m[1]-2)$ \\
$m[1][m[1]-1] + m[2][1] - u[m[1]-1] = 0$ \\
$m[1][m[1]] + m[2][2] - u[m[1]] = 0$ \\
$m[2][i+2] - u[m[1]+i] = 0$, for $i=1\ldots(m[2]-2)$ \\

For example, the function f will return True for: $m$ = \{\{4,1,2\}, \{5,2,0,1\}\}, $u$ = \{4,6,4,0,1\}. The above function is a set of equations, which can also be viewed as one ``multi-index'' equation (by analogy to the matrix equation). Now Math-Solver has to solve it with respect to the multi-index $u$, which is quite easy. Here is the solution:\\

\noindent $m[1] \geqslant 1$ \\
$m[2] \geqslant 1$ \\
$u = m[1] + m[2] - 2$ \\
$u[i] = m[1][i]$, for $i=1\ldots(m[1]-2)$ \\
$u[m[1]-1] = m[1][m[1]-1] + m[2][1]$ \\
$u[m[1]] = m[1][m[1]] + m[2][2]$ \\
$u[m[1]+i] = m[2][i+2]$, for $i=1\ldots(m[2]-2)$ \\

In fact, similarly as in Section \ref{sec:inclsol_mathsol}, we are not interested in the solution itself, but in the area in which the solution definitely exists. As can be seen, this area is given by the following inequalities:\\

\noindent $m[1] \geqslant 1$ \\
$m[2] \geqslant 1$ \\

This allows us to reduce the ($\alpha\cdot$c)$\cdot$($\alpha\cdot$c) scheme into c$\cdot$($\alpha\cdot$c)$|$($\alpha\cdot$c)$\cdot$c. This in turn allows us to reduce the ($\alpha\cdot$c)$^*$ scheme into c$^*\cdot\alpha\cdot$c$^*$. 
So, the ($\alpha\cdot$c)$^*\cdot\alpha$ scheme is therefore reduced into c$^*\cdot\alpha\cdot$c$^*\cdot\alpha$. Eventually, the latter scheme can be reduced into c$^*\cdot\alpha\cdot$c$^*$ (in an analogous, although a bit more complicated manner, as the one described above).

\subsection{Order of axioms}
\label{sec:order_azioms}
The method of generating a procedure corresponding to a given set of axioms defined at the end of Section \ref{sec:iter_scheme} is based on their arbitrary order. Naturally, this raises the question: how much a different order of axioms can affect whether we can generate a polynomial cost procedure? The answer is not simple.

There are many premises for the claim that if such a procedure can be generated, it can be generated for any order of axioms. This is the case in most simple examples. An exception is the example from Section \ref{sec:sigma_solver}, but in this case the problem lies in the imperfection of the current program. In particular, in the limitation that it can solve multi-index equations only in a naive way going hierarchically from the top (first the length of the entire multi-index, then the length of its individual elements, etc.). Eventually, in this case it is possible to obtain a good procedure also for the reverse order of axioms (although its form will be very unnatural from the human intuition point of view).

Temporarily, so that we do not have to generate procedures for all permutations of axioms, we can propose a simple algorithm: we determine the order of axioms by testing for each pair of axioms whether their order is arbitrary. In this way, we should get a relationship of weak order (if not, the problem is really difficult), which can be easily expanded to a strong order.

\section{Conclusions}
\label{sec:future_work}
The point at which we are right now can be described as the end of the beginning. What's next?

Let's return to the final algorithm described in Section \ref{sec:final_algo}. The idea of iterating through the set of axioms described at the beginning of Section \ref{sec:iter_scheme} concerns for now only Step 1. of the algorithm, and thus only the repetitive component is analyzed iteratively. It would be better to analyze as well the unambiguous component.

In the example from Section \ref{sec:more_complex} we showed how to reduce the iterative scheme to c*$\cdot\alpha\cdot$c*, which, however, is not yet the best we can do, because this scheme contains repetitions. They could be eliminated by replacing the first c* by c$^n$, where $n$ is the index of the last element in c* that can be obtained for a~given start tree. But that just needs the improvement mentioned above.

The next challenge will be to unravel the set of axioms that we will obtain by changing the axiom a in the example from Section \ref{sec:more_complex} to: \\

\noindent a: P(R(x, z), R(y, t))$\rightarrow$P(R(x, t), R(y, z)) \\
b: P(R(x, z), y)$\rightarrow$P(x, R(y, z)) \\
c: P(x, R(y, z))$\rightarrow$P(R(x, z), y) \\

This way we will get a~kind of sorting: starting from the tree t from Section \ref{sec:sigma_solver}, we can get a~tree with any permutation of the right arguments of the R nodes. However, dealing with this problem will require in addition to the above improvement also the EqualsLL operator mentioned in Section \ref{sec:split_atoms}.

The next goal, which seems to be feasible, is to make that our system can handle practically any elementary problem of TPC, where an elementary problem is one in which the number of axioms does not exceed four.

Further progress is rather the area of speculation, but it is worth to mention one more idea. For now, an unbreakable barrier are theories for which a~polynomial cost algorithm (or any with the stop property) does not exist. This, however, does not mean that nothing can be done. It could possible to partially solve such a theory by determining a procedure that will deal with a certain (syntactically determinate) subset of all theorems. For example: it is impossible to analytically solve all algebraic equations (in particular those with a degree greater than five), but all systems of linear equations can be solved.

\bibliographystyle{plain}

\end{document}